\documentstyle[12pt,titlepage]{article}
\newcommand{\be }{\begin{equation}}
\newcommand{\ee}{\end{equation}}
\newcommand{\ba}{\begin{eqnarray}}
\newcommand{\ea}{\end{eqnarray}}

\newcommand{\om}{\omega}

\newcommand{\non}{\nonumber}  

\begin{document}

\begin{titlepage}
\vspace*{-1.1truecm}

\hfill{ hep-th 9701126}

\vspace{1.7cm}

\begin{center}

{\LARGE  Charged Open Membrane Solution On A \\
Manifold With A Boundary}

\vspace{1.2cm}

{\large Fermin ALDABE$^1$ and Arne L. LARSEN$^2$}

\vspace{.5cm}

January 24, 1997

\vspace{.7cm}

{\em 
$^1$Theoretical Physics Institute, Department of Physics,\\
University of Alberta, Edmonton, Canada T6E 2J1\\
email: faldabe@phys.ualberta.ca}

\vspace{.7cm}

{\em 
$^2$Observatoire de Paris, DEMIRM, 61, Av. de l'Observatoire,\\
75014 Paris, France\\
email: larsen@mesioa.obspm.fr}

\vskip 10pt
\begin{quotation}
\baselineskip=1.0em

\vspace{1cm}

{\normalsize

{\bf Abstract}. 
Explicit
open single and multi-membrane solutions 
of  the low energy limit
of  M-theory  on the
orbifold $R^{10}\times S^1/Z_2$ are presented. 
This 
low energy action is described by an 11-dimensional  supergravity action
coupled to two $E_8$
super Yang-Mills fields, which propagate only on the 10-dimensional
boundaries of the target space.
The membrane solutions we construct preserve half the supersymmetries.  They
carry electric charge and current with 
respect to the gauge fields, whose generators are in the Cartan subalgebra of
the two $E_8$ gauge groups present at the boundaries.

 }

\end{quotation}
\end{center}

\end{titlepage}

\pagebreak
\section{Introduction}
M-theory, whose formulation is still unknown, is described
in the  low energy limit by 11-dimensional supergravity.
In addition, we know that M-theory possesses membrane solutions.
They appear as solutions of the equations of motion of the
low energy limit of M-theory.  Such solutions, in the standard case where
the 11-dimensional target space has no boundaries, have been constructed in
\cite{DS}.  
In order for these solutions to be consistent,
one must couple the 11-dimensional supergravity action to a membrane action.
This will guarantee that the solution exists everywhere. 
These solutions can describe single or multi-membrane 
configurations, and
are constructed in much the same way as single or multi-string solutions
in string theories \cite{DGHR}. However,  although charged (with respect to 
gauge fields) macroscopic string
solutions where constructed by Sen \cite{Sen} using the 
techniques of \cite{HS}, there are no known charged
membrane solutions in 11-dimensional target spaces with boundaries.

In \cite{HW2}, 11-dimensional supergravity on a target space 
with 10-dimensional
boundaries was considered.   In particular, it was 
argued that 10-dimensional
gravitational anomalies appear at the boundaries.  In order to cancel these
anomalies, it is necessary to introduce 248 vector multiplets in the adjoint 
representation of $E_8$
propagating
on  each boundary, to cancel both gravitational and gauge anomalies (the latter
arising from the presence of the 496 vector multiplets).
Then the low energy effective action, which is anomaly free, is 11-dimensional
supergravity coupled to 10-dimensional super Yang-Mills.

In \cite{HW1}, it was suggested that open membrane solutions should exists
in 11-dimensional supergravity on such a target space with 10-dimensional
boundaries.
The important requirement that open membranes must end on boundaries \cite{S},
something that is not required of open strings, is readily satisfied by
demanding that the open membrane solutions stretch between the two
10-dimensional boundaries.
The source of the
open membrane solution is described by a 3-dimensional super-membrane 
action.  At the
boundaries, this action becomes 2-dimensional and therefore allows for
2-dimensional gravitational anomalies.  In order to cancel such anomalies, it
is necessary to introduce current algebras with central charge $c=8$ at each
boundary.  Thus the anomaly free action for the super-membrane has two terms;
the first term describes the propagation of the membrane on the target space,
while the second term couples the boundaries of the membrane to $c=8$ current
algebras which are believed to generate, in the low energy limit,
the 496 vector multiplets.

The action for the current algebra can always be written, after making use
of Wilson lines, in terms of the Cartan subalgebra of each $E_8$ gauge
group.  It will be this form of the current algebra
that will be most important to us.
In finding open membrane solutions, we must consider the 11-dimensional
supergravity action with a target space having two 10-dimensional boundaries.
This action must then couple to the open membrane action described in the
previous paragraph.  This will guarantee that 
the membrane solution exists everywhere.
Our purpose in this paper is to present explicit
open membrane solutions which are charged with
respect to the gauge fields, whose generators are in the Cartan subalgebra of
the two $E_8$ gauge groups.
The paper is divided as follows.  In Section 2 we present the
action, which will support charged open
membrane solutions, and we give our conventions. 
In Section 3 we present our anzats and verify that
half the 
supersymmetries are preserved.  We show that
the equations of motion obtained from the action
are satisfied for our charged open membrane solution. 
We also construct a charged multi-open-membrane solution.
In Section 4 we give our concluding remarks. We discuss the type IA macroscopic
string obtained from dimensional reduction of the open membrane solution, and we
comment on the relation of our membrane
solution to the uncharged membrane of Duff and Stelle \cite{DS} and to
the charged macroscopic string
solution found by Sen \cite{Sen}.

\section{Action and Conventions}
The bosonic sector of 11-dimensional supergravity on $R^{10}\times 
S^1/Z_2$ is \cite{HW2}
\ba
S_{11}&=&-\frac{1}{\kappa^2}\int_{M^{11}} d^{11}x\ \sqrt{-g}\;
\Big( -\frac{1}{2} R+\frac{1}{48}
G_{MNOP}G^{MNOP}\Big)\\
 & &-\frac{\sqrt{2}}{3456\kappa^2}
\int_{M^{11}} G\wedge G\wedge C 
-\frac{1}{4\lambda^2}\int_{\partial M^{11}} 
d^{10}x\ \sqrt{-h}\; tr \Big(F^I_{AB}F^{IAB}\Big),
\nonumber
\label{11sg}
\ea
where $\partial M^{11}$ is a copy of $M^{10}$ for each boundary. 
In the first term  of (1),
$g$ is the 11-dimensional metric and $C$ is the 3-form potential whose
4-form field strength is $G$.  
In the last term, $A^I\;(I=1,...,248)$ 
are the gauge potentials filling the adjoint of
$E_8,$ whose 2-form
field strengths  are denoted by $F^I$.   The metric $h$ in the last term,
which couples to the gauge fields, is the restriction of the 11-dimensional
metric to each boundary, defined by each of the
fixed points of the interval $I\simeq 
S^1/Z_2$. We thus use indices $(M,N,O,P,..)$ 
for the 11-dimensional 
metric but indices $(A,B,C,D,..)$ for the 10-dimensional metric, and
our notations and conventions generally follow  \cite{HW2}. In particular, 
the 4-form field strength is given by
\ba
G_{zABC}&=&(\partial_{z}C_{ABC}\;\pm\; 23\; \mbox{terms})+
\frac{\kappa^2}{\sqrt{2}\lambda^2}\delta(z)\omega_{ABC},\nonumber\\
G_{ABCD}&=&\partial_AC_{BCD}\;\pm\; 23\; \mbox{terms},
\ea
where $z=x^{11}$ and $\omega_{ABC}$ is the Chern-Simons term
\be
\omega_{ABC}=tr\Big(A_A F_{BC}+\frac{2}{3}A_A[A_B,A_C]\;+\;\mbox{cyclic 
permutations}\Big).
\ee
The second term in $G_{zABC}$ is included to solve the modified Bianchi
identity, as explained in \cite{HW2}.

Concerning the gauge potentials, we introduce Wilson lines and
break each $E_8$ down to $U(1)^8,$ and we will only consider this gauge group
for the remainder of the article.
The open membrane action which couples to (1) is 
then
\ba
S_{{om}}&=&S_{{m}}
+\mu\int_{\partial M^3}d^2\zeta \epsilon^{ij}\partial_i Y^I A^I_A\partial_j X^A,
\label{om}
\ea
where
\be
S_{{m}}=-T\int_{M^3} d^3\zeta \Big(\sqrt{-\gamma}-\sqrt{2}
\epsilon^{ijk}\partial_i X^M\partial_j X^N\partial_k X^P C_{MNP}
\Big).\label{s}
\ee
In (4), $\partial M^3$ is a copy of $S^1\times R$ for each boundary. 
Also, there are 8 internal coordinates $Y^I$ at each boundary, 
obtained by bosonization 
of the left-handed Majorana fermions generating the $E_8$ current algebra, 
fulfilling
\be
\Big(\sqrt{-\eta}\;\eta^{ij}-\epsilon^{ij}\Big)\partial_i Y^{I}=0,
\ee
where the metric $\eta_{ij}$ is the restriction
of the 3-D induced metric $\gamma_{ij}$ on the world-volume of the membrane,
to the 2-D boundary of the membrane.

The action we shall use to construct the open membrane solutions is
\be
S=S_{11}+S_{om}.
\ee

\section{The Charged Open Membrane Solution} 
We introduce the following ansatz for the metric, 
\ba
ds^2&=&-(e^{2A(r)}-2 B(z,r) e^{A(r)})dt^2+
(e^{2A(r)}+ 2 B(z,r) e^{A(r)})dx^2\\
&+&4 B(z,r)e^{A(r)} dx dt
+e^{2A(r)}dz^2+e^{2Q(r)}\delta_{nm}dx^n dx^m,\nonumber
\label{ma}
\ea
and three-form potential
\be
C_{txz}= e^{C(r)}.
\label{p3a}
\ee
For the gauge 
potentials we use the
ansatz 
\be
A^I_t=A^I_x= \alpha^I e^{C(r)};\;\;\;\;\;\;\;\;
\label{p1a}
\ee
where the $\alpha^I$'s are constants.
Here the index $t$ labels the 11-dimensional timelike coordinate, $x$ is 
a spacelike  coordinate  along the membrane and
parallel to the boundary, $z$ is the spacelike coordinate also along the 
membrane, but transverse to the  boundary (already introduced in (2)).  
The remaining eight 
coordinates are labeled by indices $(m,n,p,..)$ and are all transverse to the
membrane but parallel to the boundary, and as usual, the radial coordinate
$r$ is defined by $r^2=\delta_{nm}x^n x^m.$  

Supersymmetry for a bosonic membrane 
requires that away from the boundaries the variation of the
gravitino vanishes, while at the boundary, the variation of both the
gravitino and the gaugino vanish.  These conditions can be summarized as follows
\ba
\delta \lambda^{I}&=&F^{I}_{AB}\Gamma^{AB}\epsilon=0,\\
\delta\psi_M&=&\left(
\partial_M+\frac{1}{4}\om_M\;^{\bar{M}\bar{N}}\Gamma_{\bar{M}\bar{N}}
+\frac{\sqrt{2}}{288}
(\Gamma^{PQRS}\;_M+8\Gamma^{PQR}\delta^S_M) G_{PQRS}\right) \epsilon=0,\nonumber
\ea
where $(\bar{M},\; \bar{N})$ are 11-D tangent space indices (flat indices). 

The first condition in (11), which requires the variation of the gaugino 
to vanish,
can be satisfied for our ansatz provided
 the  spinor $\epsilon$ satisfies
\be
(\Gamma^t+ \Gamma^x)\epsilon=0\label{c1}
\ee
at the boundary.
The non vanishing elfbeins  corresponding to the metric (8) are
\ba
&&E_t^{\bar{t}}=-e^A+B,\;\;\;\;\;E_x^{\bar{x}}=e^A+B,\;\;\;\;\;
E_x^{\bar{t}}=E_t^{\bar{x}}=B,\non\\
&&E_m^{\bar{n}}=\delta^{\bar{n}}_m e^Q,\;\;\;\;\;E_z^{\bar{z}}=e^A.
\ea
Then, from (12), it follows that
\be
(\Gamma_{\bar{t}}+\Gamma_{\bar{x}})\epsilon=0.
\ee
This condition is equivalent to requiring that the spinor be chiral
in the 10-D sense,
and we have the same situation as encountered by Sen \cite{Sen}.

Next we consider the gravitino variations. Following Duff and Stelle
\cite{DS}, we decompose the gamma-matrices in the following way
\be
\Gamma_{\bar{M}}=(\gamma_{\bar{\alpha}}\otimes \Gamma_9, 
1\otimes\Sigma_{\bar{a}}),
\ee
where $\gamma_{\bar{\alpha}}$ and $\Sigma_{\bar{a}}$ 
are the D=3 and D=8 (flat) Dirac 
matrices, respectively, while $\Gamma_9\equiv\Sigma_{\bar{4}}
\Sigma_{\bar{5}}...\Sigma_{\bar{11}}$, where the bar denotes 
tangent space indices (flat indices).

We shall also use extensively the picture that spacetime 
$R^{10}\times S^1/Z_2$ can be considered a manifold with boundaries.
In the following we consider for simplicity 
only one of the boundaries. Let it be defined
by $z=0,$ and let the bulk-part of spacetime correspond to $z<0$. The
function $B(z,r),$ as introduced in the metric (8), is now decomposed 
as
follows
\be
B(z,r)=f(r)(z+|z|),
\ee
so that $B(z,r)$ vanishes everywhere in bulk and at the boundary, but its
$z$-derivatives survive at the boundary. Now using (16), as well as (12) 
at the boundary, 
the  gravitino 
variations 
reduce for $z\leq 0$ to
\ba
\delta\psi_t&=&\Big(\partial_t+\frac{1}{2}
\gamma_t \Sigma^m A_{,m}\Gamma_9
+\sqrt{2}\gamma_t e^{-3A+C}\Sigma^m C_{,m}\Big)\epsilon,\non\\
\delta\psi_x&=&\Big(\partial_x+\frac{1}{2}
\gamma_x \Sigma^m A_{,m}\Gamma_9
+ \sqrt{2}\gamma_x e^{-3A+C}\Sigma^m C_{,m}\Big)\epsilon,\\
\delta\psi_z&=&\Big(\partial_z+\frac{1}{2}
\gamma_z \Sigma^m A_{,m}\Gamma_9
+ \sqrt{2}\gamma_z e^{-3A+C}\Sigma^m C_{,m}\Big)\epsilon,\non\\
\delta\psi_m&=&
\Big(\partial_m-\frac{1}{4}(\Sigma_m\Sigma^n-\Sigma^n\Sigma_m)
Q_{,n}-\sqrt{2} e^{-3A+C} C_{,m}\Gamma_9\non\\
&+& \frac{\sqrt{2}}{4} e^{-3A+C} (\Sigma_m\Sigma^n-\Sigma^n\Sigma_m) 
 C_{,n}\Gamma_9\Big)\epsilon.\non
\ea
Notice that we have skipped terms automatically 
vanishing for $z\leq 0$ (but generally not
vanishing outside spacetime, corresponding to 
$z>0$) due to (16), as well as terms multiplying
$(\Gamma^t+\Gamma^x)$, c.f. (12).
Then if
\ba
C&=&3A-\log(6\sqrt{2}),\non\\
Q&=&-A/2,
\ea
and provided that
\be
\epsilon=e^{-A/2}\epsilon_o,
\ee
where the constant spinor $\epsilon_o$ satisfies
\be
(1+\Gamma_9)\epsilon_o=0,
\ee
 we find that all supersymmetry variations of
the gravitino and gaugino vanish in bulk ($z<0$) and at 
the boundary ($z=0$).  Notice that the two conditions 
$(1+\Gamma_9)\epsilon=0$ and 
$(\Gamma^t+\Gamma^x)\epsilon=0$ are 
equivalent since\footnote{We thank P. Horava for a clarifying comment to us
about this point.}
\be
\Gamma_{\bar{z}}\epsilon=\epsilon\;\;\;\;\;\;\mbox{and}\;\;\;\;\;\;
\Gamma_9\Gamma_{\bar{t}}\Gamma_{\bar{x}}\Gamma_{\bar{z}}=-1
\ee
on 
$R^{10}\times S^1/Z_2,\;$
\cite{HW1}. The conditions obtained for the spinor thus
correspond to breaking half the supersymmetries in bulk and at the boundary.

Using the following embedding for the membrane
\be
X^{\mu}(\zeta)=\zeta^{\mu},\ \mu=t,x,z\;\;\; \mbox{and}\;\; X^m(\zeta)=
\mbox{const.},\label{f3}
\ee
the 
equation of motion for the G-field, for our ansatz, is
\be
\partial_m (\sqrt{-g}\;G^{mtxz})
 =-\sqrt{2}T\kappa^2 \delta^8(r),\label{e2}
\ee
taking the location of the core of the membrane to be at $r=0.$
This equation is solved by

\be
A(r)=-\frac{1}{3}\log(1+\frac{K}{r^6}),
\ee
provided the constant $K$ is related to the membrane tension by
\be
K=\frac{T\kappa^2}{3\Omega_8},
\ee
where $\Omega_8$ is the area of the unit sphere in $R^8.$

In addition we find, for our ansatz and the embedding (22), that 
the equations of motion for the gauge fields are
\be
\partial_n(\sqrt{-h}F^{I n\mu})+\sqrt{2}\sqrt{-g}G^{n\nu\mu z}F^I_{n\nu}=
\mu\lambda^2\epsilon^{\rho\nu} \partial_\rho Y^I\partial_\nu X^\mu\delta^{8}(r),
\ee
where $\mu,\nu,\rho,..$ are either $t$ or $x$. Notice that the 
gauge-dependent term, which arises when varying the Chern-Simons term,
has been omitted, c.f. \cite{Sen}. 
This equation is solved by (24), (10) and (9) provided
\be
Y^I=\frac{K\alpha^I\Omega_8}{\sqrt{2}\mu\lambda^2}e^{3A}(t+x),
\ee
where the function $A$ is evaluated at $X^m$=const., c.f. (22). Notice
also that (27) is consistent with (6), as it should be.

Next we consider the Einstein equations
\be
R_{MN}-\frac{1}{2} g_{MN} R=T_{MN}.
\ee
For completeness we list below the non-vanishing components of the 
Ricci-tensor and the energy-momentum tensor 
\ba
R_{zz}&=&-e^{3A}\delta^{nm}A_{,nm},\nonumber\\
R_{nm}&=&\frac{1}{2}\delta_{nm}\delta^{pq}A_{,pq}-
\frac{9}{2}A_{,n}A_{,m},\\
\left( \begin{array}{cc} R_{tt} & R_{tx} \\  R_{xt} & R_{xx}
\end{array}\right)&=&-e^{3A}\delta^{nm}A_{,nm}
\left( \begin{array}{cc} -1 & 0 \\  0 & 1
\end{array}\right)-e^{-A}B''
\left( \begin{array}{cc} 1 & 1 \\  1 & 1
\end{array}\right),\nonumber
\ea
so that 
\be
R=e^A(\delta^{nm}A_{,nm}-\frac{9}{2}\delta^{nm}A_{,n}A_{,m}).
\ee
The energy-momentum tensor takes the form
\ba
T_{zz}&=&-\frac{9}{4}e^{3A}\delta^{nm}A_{,n}A_{,m}-\kappa^2Te^{6A}
\delta^8(r),\nonumber\\
T_{nm}&=&-\frac{9}{2}A_{,n}A_{,m}+\frac{9}{4}\delta_{nm}
\delta^{pq}A_{,p}A_{,q},\\
\left( \begin{array}{cc} T_{tt} & T_{tx} \\  T_{xt} & T_{xx}
\end{array}\right)&=&-\Big(\frac{9}{4}e^{3A}\delta^{nm}A_{,n}A_{,m}+
\kappa^2Te^{6A}
\delta^8(r)\Big)
\left( \begin{array}{cc} -1 & 0 \\  0 & 1
\end{array}\right)\nonumber\\
&+&\frac{\kappa^2}{\lambda^2}\delta(z)
\delta^{nm}A^I_{t,n}A^I_{t,m}
\left( \begin{array}{cc} 1 & 1 \\  1 & 1
\end{array}\right),\nonumber    
\ea
where we used that $A^I_t=A^I_x$. Again it should be stressed that we have 
skipped terms automatically vanishing for $z\leq 0$ (but generally not
vanishing for $z>0$). Notice also that the $z$-dependence in the 
energy-momentum tensor appears only via the delta-function multiplying the 
Yang-Mills term. This delta-function arises because the gauge fields live only
at the boundaries. Similarly, the $z$-dependence in the Einstein tensor
appears only in the $t-x$ sector due to the $z$-dependence of the metric (8) 
in the $t-x$ sector.

It is now straightforward to check that all
the Einstein equations are consistently fulfilled by our ansatz,
provided the function $f(r),$ as introduced in (16), is given by
\be
f(r)=\frac{-\kappa^2}{2\lambda^2}\delta^{nm}A^I_{t,n}A^I_{t,m}e^A,
\ee
where $A^I_{t}$ is given by (10), (18) and (24).

Finally we consider the equations of motion for the membrane source, as
obtained from (4), (5). For our ansatz and embedding, the boundary term in
(4) does not give any contribution so we are left with
\ba
\frac{1}{2}\sqrt{-\gamma}\gamma^{ij}X^M_{,i}X^N_{,j}g_{MN,P}
&-&\partial_j(\sqrt{-\gamma}\gamma^{ij}g_{MP}X^M_{,i})\\
&=&-\frac{\sqrt{2}}{6}
\epsilon^{ijk}\partial_i X^M\partial_j X^N\partial_k X^O
G_{PMNO}.\nonumber
\ea
Again we find that our ansats and embedding exactly fulfills the
equations for $z\leq 0$. This concludes the proof that our open membrane
with boundary solves all the equations for $g_{MN},\;C_{MNP},\;A_{A}$
as well as the equations for the membrane source itself.

Then, the  open membrane solution  is
\ba
ds^2&=&
-\left[ \left(
1+\frac{K}{r^6}\right)^{-\frac{2}{3}}+
\frac{\kappa^2(\alpha^I)^2 K^2}{2\lambda^2 r^{14}}\; (z+|z|)\
 \left(1+\frac{K}{r^6}\right)^{-\frac{14}{3}}
\right]dt^2
\non\\
&&+\left[ \left(
1+\frac{K}{r^6}\right)^{-\frac{2}{3}}-
\frac{\kappa^2(\alpha^I)^2 K^2}{2\lambda^2 r^{14}}\; (z+|z|)\
 \left(1+\frac{K}{r^6}\right)^{-\frac{14}{3}}
\right]dx^2
\non\\
&&-\left[ 
\frac{\kappa^2(\alpha^I)^2 K^2}{\lambda^2 r^{14}}\; (z+|z|)\
 \left(1+\frac{K}{r^6}\right)^{-\frac{14}{3}}
\right]dx\ dt
\non\\
&& +\left(1+\frac{K}{r^6}\right)^{-\frac{2}{3}}dz^2+
\left(1+\frac{K}{r^6}\right)^\frac{1}{3}
\delta_{nm}dx^n dx^m,
\non\\
A^I_t&=&A^I_x=\frac{\alpha^I}{6\sqrt{2}} \left(1+\frac{K}{r^6}\right)^{-1}
\non\\
C_{txz}&=&
\frac{1}{6\sqrt{2}} \left(1+\frac{K}{r^6}\right)^{-1}.\label{sol23}
\ea

Before carrying on, we should pause to make a statement about the parity
of the membrane solution (\ref{sol23}).
It is sometimes convenient to consider the target space to be 
$S^1\times R^{10}$ instead of $S^1/Z_2\times R^{10}$.  
We should then impose parity-invariance around some point 
$z\in S^1,\;$ (say) $z=0.$  Under a parity
transformation
\be
z\to-z,
\ee
we must demand that 
\ba
C_{ABC}(z)&=&-C_{ABC}(-z),\non\\
C_{ABz}(z)&=& C_{ABz}(-z),\non\\
g_{AB}(z)&=&g_{AB}(-z),\non\\
g_{Az}(z)&=&-g_{Az}(-z),\non\\
g_{zz}(z)&=&g_{zz}(-z),\label{do9}
\ea
where the indices $(A, B, C)$ are 10-dimensional.
The requirements (\ref{do9}), are equivalent to demanding that the solutions
on the target space
 $S^1\times R^{10}$ also be solutions on the target space
$S^1/Z_2\times R^{10}$.  It is clear from our anzats, that the metric
does not satisfy all the conditions (\ref{do9}).  However, 
in order to avoid this 
inconsistency, we should use the ansatz (16) only when the 
spacetime is identified with
$z\leq 0.$ If spacetime is instead 
identified with $z\geq 0,$ we should use instead
of (16) the ansatz $B(z,r)=f(r)(-z+|z|).$ 
In this way, the metric $g_{AB}$ actually is, by definition, even
under $z\to -z$, and the only effect of the "$|z|$" is to produce 
a non-vanishing $z$-derivative at $z=0,$ which is necessary in the Einstein
tensor, as explained after equation (31). However, for our purposes here it is
generally most convenient to actually think of an 11-dimensional spacetime with 
boundaries.

We must now show that the 
membrane source  (\ref{f3})
also preserves half the supersymmetries.
Following \cite{berg}, we must require for a bosonic membrane source that 
the Killing spinor also satisfies
\be
(1+\Gamma)\epsilon=0\label{co1}
\ee
where 
\be
\Gamma \equiv \frac{1}{3!\sqrt{-\gamma}}\epsilon^{ijk}\partial_iX^M
\partial_jX^N
\partial_kX^P\Gamma_{MNP}.
\ee
 Using (15), we find that $\Gamma=-\Gamma_{\bar{t}}\Gamma_{\bar{x}}
\Gamma_{\bar{z}}$ and then using (21), the condition (37) reduces to  
\be
(1+\Gamma_9)\epsilon=0,
\ee
that is, equation (20).
Hence we are able to satisfy (\ref{co1}) in the same manner as done
in \cite{DS}. At the boundaries we should strictly speaking use instead of
(38) the expression
\be
\Gamma \equiv \frac{1}{2!\sqrt{-\eta}}\epsilon^{ij}\partial_iX^M
\partial_jX^N
\Gamma_{MN}.
\ee     
Then we get $\Gamma=-\Gamma_{\bar{t}}\Gamma_{\bar{x}},$
which from (21) again reduces to (39) when acting on the spinor.

The charges and currents of the membrane solution can be read off 
from the 2-form
field strength and the 4-form field strength in the limit that $r\to\infty$
\ba
F^I_{xr}&=&F^I_{tr}=\frac{K\alpha^I}{\sqrt{2}r^7}\Big(1+\frac{K}{r^6}\Big)^{-2}
\;\sim\;\frac{K\alpha^I}{\sqrt{2}r^7}\;,\non\\
G_{txzr}&=&\frac{6K}{\sqrt{2}r^7}\Big(1+\frac{K}{r^6}\Big)^{-2}
\;\sim\;\frac{6K}{\sqrt{2}r^7}\;.
\ea
Using the standard formula for a 1-brane in 10 dimensions, we conclude that
the $U(1)$-charges and currents at the boundary are (up to a numerical factor)
\be
Q^I=J^I=K\alpha^I.
\ee
Similarly, using the standard formula for a 2-brane in 11 dimensions, we 
conclude that
the "axion-charge" is (up to a numerical factor) 
\be
Z=K.
\ee
Although we have only constructed a single  open membrane solution, it
is straightforward to show
that a stable multi-membrane solution is obtained by a linear
superposition of solutions
\be
A(\vec{r})=-\frac{1}{3}\log\left( 1+
\sum_i\frac{K}{|\vec{r}-\vec{r}_i|^6}\right),\label{f4}
\ee
where $\vec{r}_i$ denotes the location of the core of the $i'th$ membrane,
together with the corresponding generalization of the embedding (22).
This is the typical situation in soliton theory, so we shall not go into 
more details here. The multi-membrane solution takes the form (34) after
the substitution
\be
\left( 1+\frac{K}{r^6}\right)\;\longrightarrow\;\left( 1+
\sum_i\frac{K}{|\vec{r}-\vec{r}_i|^6}\right).
\ee
that is
\ba
ds^2&=&
-\left[ 
\left( 1+\sum_i\frac{K}{|\vec{r}-\vec{r_i}|^6}\right)
^{-\frac{2}{3}}+
\frac{\kappa^2(\alpha^I)^2 K^2}{2\lambda^2 r^{14}}\ (z+|z|)\
\left( 1+\sum_i\frac{K}{|\vec{r}-\vec{r_i}|^6}\right)
^{-\frac{14}{3}}
\right]dt^2
\non\\
&&+\left[ 
\left( 1+\sum_i\frac{K}{|\vec{r}-\vec{r_i}|^6}\right)
^{-\frac{2}{3}}-
\frac{\kappa^2(\alpha^I)^2 K^2}{2\lambda^2 r^{14}}\ (z+|z|)\
\left( 1+\sum_i\frac{K}{|\vec{r}-\vec{r_i}|^6}\right)
^{-\frac{14}{3}}
\right]dx^2
\non\\
&&-\left[ 
\frac{\kappa^2(\alpha^I)^2 K^2}{\lambda^2 r^{14}}\ (z+|z|)\
\left( 1+\sum_i\frac{K}{|\vec{r}-\vec{r_i}|^6}\right)
^{-\frac{14}{3}}
\right]dx\ dt
\non\\
&& +
\left( 1+\sum_i\frac{K}{|\vec{r}-\vec{r_i}|^6}\right)
^{-\frac{2}{3}}dz^2+
\left( 1+\sum_i\frac{K}{|\vec{r}-\vec{r_i}|^6}\right)
^\frac{1}{3}
\delta_{nm}dx^n dx^m,
\non\\
A^I_t&=&A^I_x=\frac{\alpha^I}{6\sqrt{2}} 
\left( 1+\sum_i\frac{K}{|\vec{r}-\vec{r_i}|^6}\right)
^{-1}
\non\\
C_{txz}&=&
\frac{1}{6\sqrt{2}} 
\left( 1+\sum_i\frac{K}{|\vec{r}-\vec{r_i}|^6}\right)
^{-1}.
\ea
and the charges and currents of each membrane in the multi-membrane solution 
equal those of the single membrane solution.

\section{ Concluding Remarks}
In conclusion, we have constructed open single and multi-membrane solutions
of  the low energy limit
of  M-theory  on the
orbifold $R^{10}\times S^1/Z_2.$ 
These membrane solutions were shown to 
preserve half the supersymmetries.  They carry
"axion-charge" in the 11-dimensional spacetime as well as
electric charge and current with
respect to the gauge fields, whose generators are in the Cartan subalgebra of
the two $E_8$ gauge groups present at the 10-dimensional boundaries.    

There are in principle 
different types of dimensional reductions that can be performed
on our membrane solution. Let us consider now the orbifold 
$R^9\times S^1\times S^1/Z_2$ and let $x$ be the coordinate on $S^1.$ 
When dimensionally reducing along the $x$-direction, the resultant supergravity 
action is that of type IA supergravity 
\cite{HW1}.  The resultant source action is
that of the type IA string  which couples to $SO(16)\times SO(16)$
Chan Paton factors \cite{A}.
It turns out that the dimensional reduction of the open membrane solution is 
the open string solution with Chan Paton factors
generated by dimensional reduction of the term which couples the 
 gauge fields to the membrane in the source term.
Thus this solution is macroscopic  type IA string solution.

When performing a standard Kaluza-Klein 
dimensional reduction along the $z$-direction, the resultant
supergravity action is the low energy limit of the heterotic string theory.
However, at the present stage, it is not clear how to perform the corresponding
dimensional reduction of our membrane solution. The reason is that the 
solution depends explicitly on the $z$-coordinate 
(even in a discontinuous way), so 
the standard ansatz for dimensional reduction does not work here.
But notice that our membrane solution shows several similarities with  
the string solution 
found by Sen \cite{Sen}, thus we expect that our solution should 
actually reduce to Sens solution after a dimensional reduction along the 
$z$-direction. Then our solution would naturally interpolate between the
membrane solution of Duff and Stelle \cite{DS} in bulk and the charged 
string solution of Sen \cite{Sen} at the boundaries.

\vskip 48pt
\hspace*{-6mm}{\bf Acknowledgements:}\\
We would like to thank H. Balasin, N. Kaloper and specially R. Myers
for helpfull discussions.
The work of F. Aldabe  was supported by NSERC, Canada.
A.L. Larsen was partly supported by NSERC, Canada and partly by CNRS, France.


\newpage
\begin{thebibliography}{11}

\bibitem{DS}M.J. Duff and K.S. Stelle, {\em Multimembrane Solutions Of
D=11 Supergravity
}, Phys. Lett. {\bf B253}
(1991) 113.   


\bibitem{DGHR}A. Dabholkar, G.W. Gibbons, J.A. Harvey and
F. Ruiz-Ruiz, {\em Superstrings And Solitons
}, Nucl.
Phys. {\bf B340} (1990) 33.

\bibitem{Sen}A. Sen, {\em Macroscopic Charged Heterotic String 
}, Nucl. Phys. {\bf B388} (1992) 457.

\bibitem{HS} S.F. Hassan and A. Sen, {\em Twisting Classical Solutions In
The Heterotic String Theory}, Nucl. Phys. {\bf B375} (1992) 103.

\bibitem{HW2}P. Horava and E. Witten,
{\em Eleven-Dimensional Supergravity
on A Manifold with Boundary}, Nucl. Phys. {\bf B475} (1996) 94.  

\bibitem{HW1}P. Horava and E. Witten, 
 {\em Heterotic and Type I String Dynamics
>From Eleven Dimensions},
Nucl. Phys. {\bf B460} (1996) 506.

\bibitem{S} A. Strominger, {\em Open p-Branes},
Phys. Lett. {\bf B383} (1996) 44.

\bibitem{berg}E. Bergshoeff, E. Sezgin and P.K. Townsend, 
{\em Supermembranes And Eleven-Dimensional Supergravity}, Phys.
Lett. {\bf B189} (1987) 75.

\bibitem{A} F. Aldabe, {\em Heterotic and Type I Strings from Twisted 
Supermembranes},
 Nucl. Phys. {\bf B473} (1996) 63.

\end{thebibliography}
\end{document}